\documentclass[preprint,10pt]{elsarticle}

\usepackage{amssymb}
\usepackage{amsmath}

\newtheorem{as}{Assumption}[section]
\newtheorem{thm}{Theorem}[section]
\newtheorem{lem}{Lemma}[section]

\makeatletter

\@addtoreset{equation}{section}

\def\be{{\beta}}
\def\ga{{\gamma}}

\def\ep{{\varepsilon}}
\def\la{{\lambda}}
\def\si{{\sigma}}
\def\th{{\theta}}

\def\bbe{{\text{\boldmath $\beta$}}}

\def\bphi{{\text{\boldmath $\phi$}}}

\def\beh{{\widehat \beta}}
\def\gah{{\widehat \ga}}
\def\thh{{\widehat \th}}

\def\sih{{\widehat \si}}

\def\muh{{\widehat \mu}}
\def\lah{{\widehat \la}}

\def\bbeh{{\widehat \bbe}}

\def\bphih{{\widehat \bphi}}

\def\mut{\widetilde{\mu}}

\def\tht{\widetilde{\theta}}

\def\0{{\text{\boldmath $0$}}}
\def\1{{\text{\boldmath $1$}}}

\def\b{{\text{\boldmath $b$}}}
\def\x{{\text{\boldmath $x$}}}

\def\I{{\text{\boldmath $I$}}}

\def\V{{\text{\boldmath $V$}}}

\def\tr{{\rm tr}}

\def\Re{\mathbb{R}}

\def\E{{\rm E}}

\def\Bh{\widehat{B}}
\def\Ah{\widehat{A}}
\def\phih{\widehat{\phi}}

\begin{document}
\begin{frontmatter}

\title{Transforming Response Values in Small Area Prediction}

\author{Shonosuke Sugasawa} 
\ead{sugasawa@ism.ac.jp}
\address{The Institute of Statistical Mathematics, 10-3 Midori-cho, Tachikawa-shi, Tokyo 190-8562, Japan.}

\author{Tatsuya Kubokawa}
\ead{tatsuya@e.u-tokyo.ac.jp}
\address{Faculty of Economics, University of Tokyo, 7-3-1 Hongo, Bunkyo-ku, Tokyo 113-0033, Japan.}

\begin{abstract}
In real applications of small area estimation, one often encounters data with positive response values.
The use of a parametric transformation for positive response values in the Fay-Herriot model is proposed for such a case.
An asymptotically unbiased small area predictor is derived and a second-order unbiased estimator of the mean squared error is established using the parametric bootstrap.
Through simulation studies, a finite sample performance of the proposed predictor and the MSE estimator is investigated.
The methodology is also successfully applied to Japanese survey data.
\end{abstract}


\end{frontmatter}

\section{Introduction}

Sample surveys are indispensable to estimate various characteristics of a population of interest. 
However, reliability of estimates from sample surveys depends on sample sizes, and direct estimates from small sample surveys have large variability, which is known as a small area estimation problem.
In small area estimation methodology, a model-based approach has become very popular to produce indirect and improved estimates by `borrowing strength' from related areas.
Importance and usefulness of the model-based small area estimation approach has been emphasized in the literature. 
For a recent comprehensive review of small area estimation, see Pfeffermann (2013) and Rao and Molina (2015).

To describe the detailed setting, we define $y_i$ as the direct survey estimator of the area mean $\th_i$, noting $y_i$ is often unstable because of small area sample sizes.
For producing a reliable estimate of $\th_i$, the most famous and basic small area model is the Fay-Herriot (FH) (Fay and Herriot, 1979) described as
\begin{equation}\label{FH}
y_i=\x_i^t\bbe+v_i+\ep_i, \ \ \ i=1,\ldots,m,
\end{equation}
where $v_i\sim N(0,A)$ and $\ep_i\sim N(0,D_i)$ for known $D_i$'s, and the quantity of interest is $\th_i=\x_i^t\bbe+v_i$.
It is well known that the best predictor $\tht_i$ that minimizes the mean squared error is expressed as
$$
\tht_i=\gamma_i y_i+(1-\gamma_i)\x_i^t\bbe,
$$
which is a weighted combination of the direct estimator $y_i$ and the synthetic part $\x_i^t\bbe$ with weight $\gamma_i=A/(A+D_i)$.
The weight is a decreasing function of $D_i$ so that the weight on synthetic part $\x_i^t\bbe$ is large when $y_i$ is not reliable, that is, the sampling variance $D_i$ is large.
Since it depends on unknown parameters $\bbe$ and $A$, the practical form of $\tht_i$ is obtained by plugging estimators $\bbeh$ and $\Ah$ into $\tht_i$, namely
$$
\thh_i=\gah_i y_i+(1-\gah_i)\x_i^t\bbeh=\frac{\Ah y_i+D_i\x_i^t\bbeh}{\Ah+D_i},
$$
which is called the empirical best linear predictor (EBLUP).

Until now, the EBLUP and the related topics have been extensively studied in the framework of the Fay-Herriot model.
Chatterjee et al. (2008) and Diao et al. (2014) proposed the empirical Bayes confidence intervals of $\th_i$ with second-order refinement.
Li and Lahiri (2010) and Yoshimori and Lahiri (2014) were concerned with the problem of estimating the variance parameter $A$ avoiding $0$ estimate.
Moreover, Ghosh et al. (2008) and Sinha and Rao (2009) suggested some robust estimating methods for the Fay-Herriot model.
The Fay-Herriot model and EBLUP are simple and useful methods, but the setting of the Fay-Herriot model is sometimes inadequate for analysis of real data.
Therefore, several extensions of the Fay-Herriot model have been proposed.
Opsomer et al. (2008) suggested a nonparametric small area model using penalized spline regression.
In relation to the assumption of known $D_i$'s, Gonz\'{a}lez-Manteiga et al. (2010) proposed a nonparametric procedure for estimating $D_i$, and You and Chapman (2006) and Maiti et al. (2014) considered shrinkage estimation of $D_i$ by assuming a hierarchical structure for $D_i$.
Ybarra and Lohr (2008) and Arima et al. (2014) were concerned with the problem of measurement error in covariate $\x_i$.
Datta et al. (2011) and Molina et al. (2015) suggested procedures of preliminary testing for existence of the random effect $v_i$.
Datta and Mandal (2015) proposed a mixture of two models; one includes a random effect and the other does not include a random effect.
Although all these papers treat important problems, the response values of the data are assumed to be normally distributed.

However, we often encounter positive-valued data (e.g. income, expense), which have skewed distributions and non-linear relationships with covariates.
For such a data set, the traditional Fay-Herriot model with a linear structure between response values and covariates and a normally distributed error term is not appropriate.
A typical alternative approach is using the log-transformed response values as discussed in Slud and Maiti (2006), but the log-transformation is not always appropriate and it may produce inefficient and biased prediction when the log-transformation is misspecified.
Thus, a natural way to solve this problem is using a parametric family of transformations which enables us to select a reasonable transformation based on data.
A famous family is the Box-Cox transformation (Box and Cox, 1964) defined as
$$
h^{BC}_{\la}(x)=\begin{cases}
\la^{-1}(x^{\la}-1)   & \la\neq 0\\
\log x & \la=0.
\end{cases}
$$
However, it suffers from a truncation problem that the range of the Box-Cox transformation is not the whole real line if $\la\neq 0$, which leads to inconsistency of the maximum likelihood estimator of $\la$.
Moreover, the inverse transformation cannot be defined on whole real line, so that we cannot define a back-transformed predictor in the original scale.
Alternatively, Yang (2006) suggested a novel family of transformations called the dual power transformation (DPT):
$$
h_{\la}(x)=\begin{cases}
(2\la)^{-1}(x^{\la}-x^{-\la})   & \la>0\\
\log x & \la=0,
\end{cases}
$$
which can be seen as the average of two Box-Cox transformations, namely $h_{\la}(x)=\{h^{BC}_{\la}(x)+h^{BC}_{-\la}(x)\}/2$.
The main advantage of the DPT is that its range is the whole real line for all $\la\geq 0$ so that DPT does not suffer from the truncation problem.
Sugasawa and Kubokawa (2015) proposed the Fay-Herriot model in which response variables are transformed by general parametric transformations.
In this paper, we focus on the FH model with DPT transformation described as
\begin{equation}\label{TFH2}
h_\la(y_i)=\x_i^t\bbe+v_i+\ep_i, \ \ \ \ i=1,\ldots,m,
\end{equation}
where $v_i\sim N(0,A)$ and $\ep_i\sim N(0,D_i)$ for known $D_i$'s.

Although Sugasawa and Kubokawa (2015) derived EBLUP of $\th_i=\x_i^t\bbe+v_i$ and the MSE estimator, the parameter of most interest in the model (\ref{TFH2}) is $\mu_i=h_{\la}^{-1}(\th_i)$ rather than $\th_i$, where $h_{\la}^{-1}(\cdot)$ is the inverse transformation of DPT:
$$
h_{\la}^{-1}(x)=\left(\la x+\sqrt{\la^2x^2+1}\right)^{1/\la}.
$$
Thus, the method developed in Sugasawa and Kubokawa (2015) is not enough for practical applications.
In this paper, we focus on the prediction of $\mu_i$ with its risk evaluation.
Specifically, we derive the best predictor of $\mu_i$ as the conditional expectation and the empirical best predictor by plugging the parameter estimates in the best predictor.
For risk evaluation, we construct a second order unbiased MSE estimator based on the parametric bootstrap.

The paper is organized as follows.
In Section \ref{sec:TFH}, we derive the best predictors of $\mu_i$ as well as the maximum likelihood estimation of model parameters.
A second-order unbiased estimator of the mean squared error of the small area predictor is derived based on the parametric bootstrap.
In Sections \ref{sec:sim} and \ref{sec:emp}, we show some simulation studies and empirical applications, respectively.
In Section \ref{sec:conc}, we give some concluding remarks. 
The technical details are given in the Appendix.

\section{Transformed Fay-Herriot Model}\label{sec:TFH}


\subsection{Model setup and best predictor}\label{sec:model}
We consider the following parametric transformed Fay-Herriot (PTFH) model for area-level data:
\begin{equation}\label{TFH}
h_\la(y_i)=\x_i^t\bbe+v_i+\ep_i, \ \ \ \ i=1,\ldots,m
\end{equation}
where $v_i\sim N(0,A)$, $\ep_i\sim N(0,D_i)$ for known $D_i$'s, $\bbe$ and $\x_i$ are $p$-dimensional vectors of regression coefficients and covariates, respectively.
The unknown models parameters are denoted by $\bphi=(\bbe^t,A,\la)^t$ and we aim to estimate (predict) $\mu_i=h^{-1}_{\la}(\th_i)$ with $\th_i=\x_i^t\bbe+v_i$.
Note that when $\la=0$, the model (\ref{TFH}) reduces to the log-transformed Fay-Herriot model studied by Slud and Maiti (2006).

It is well known that the best predictor of $\th_i$ under the squared error loss is given by
\begin{equation}\label{tht}
\tht_i=\ga_i h_{\la}(y_i)+(1-\ga_i)\x_i^t\bbe,
\end{equation}
where $\ga_i=A/(A+D_i)$.
Hence, one possible way to predict $\mu_i$ is using the simple back-transformed predictor $\mut_i^{(S)}=h_{\la}^{-1}(\tht_i)$.
However,  $\mut_i^{(S)}$ is not suitable for predicting $\mu_i$, because $\mut_i^{(S)}$ has a non-ignorable bias for predicting $\mu_i$, namely $\E[\mut_i^{(S)}-\mu_i]\neq 0$ even when $m$ is large.
On the other hand, Slud and Maiti (2006) considered the bias corrected predictor of $F(\th_i)$ for a general function $F(\cdot)$, which leads to the following form:
$$
\mut_i^{(SM)}=\frac{\E[h_{\la}^{-1}(\th_i)]}{\E[h_{\la}^{-1}(\tht_i)]}h_{\la}(\tht_i)=\frac{\E[\mu_i]}{\E[\mut_i^{(S)}]}\mut_i^{(S)}.
$$
It clearly holds that $\E[\mut_i^{(SM)}-\mu_i]=0$, that is, $\mut_i^{(SM)}$ is an unbiased predictor of $\mu_i$ while it does not necessarily minimize the squared error loss.
We here use the conditional expectation $\mut_i=\E[h_{\la}^{-1}(\th_i)|y_i]$ with known $\bphi$ as a predictor of $\mu_i$, which minimizes the squared error loss.
Since $\th_i|y_i\sim N(\tht_i,\si_i^2)$ with $\tht_i$ given in (\ref{tht}) and $\si_i^2=AD_i/(A+D_i)$ under the model (\ref{TFH}), the conditional expectation $\mut_i$ can be expressed as 
\begin{equation}\label{BP}
\mut_i\equiv \mut_i(y_i;\bphi)=\int_{-\infty}^{\infty}h_{\la}^{-1}(t)\phi(t; \tht_i,\si_i^2)dt,
\end{equation}
where $\phi(\cdot; a,b)$ denotes the density function of $N(a,b)$.
It should be noted that $\mut_i=\mut_i^{(SM)}$ when $\la=0$, namely $h_{\la}^{-1}(x)=\exp(x)$.
However, $\mut_i$ and $\mut_i^{(SM)}$ are not necessarily identical when $\la>0$.

Since the model parameters $\bphi$ is unknown in practice, we estimate them by maximizing the marginal likelihood function, and the details are given in the next section.
Let $\bphih$ be the corresponding estimator of $\bphi$.
Then, replacing $\bphi$ with $\bphih$ in (\ref{BP}) leads to the empirical form of $\mut_i$:
$$
\muh_i\equiv \mut(y_i;\bphih)=\int_{-\infty}^{\infty}h_{\lah}^{-1}(t)\phi(t; \thh_i,\sih_i^2)dt,
$$
which is known as empirical best predictor (EBP).
Note that $\muh_i$ is no longer the conditional expectation but $\muh_i$ converges to $\mut_i$ as $m\to\infty$ under some regularity conditions.
Since $\muh_i$ cannot be obtained in an analytical form, we rely on numerical techniques for computing $\muh_i$.
A typical method is the Monte Carlo integration by generating a large numbers of random samples from $(\thh_i,\sih_i^2)$.
However, we here use Gaussian-Hermite quadrature which is known to be more accurate than the Monte Carlo integration.


\subsection{Estimation of model parameters}\label{sec:est}
Under normality assumption of $v_i$ and $\ep_i$, it follows that $h_{\la}(y_i)\sim N(\x_i^t\bbe,A+D_i)$ and $h_{\la}(y_i),\ i=1,\ldots,m$ are mutually independent.
Then, the maximum likelihood estimator $\bphih$ of $\bphi$ is defined as the maximizer of $L(\bphi)$, where
\begin{equation}\label{like}
L(\bphi)=-\sum_{i=1}^m\log(A+D_i)-\sum_{i=1}^m\frac{\left\{h_{\la}(y_i)-\x_i^t\bbe\right\}^2}{A+D_i}+2\sum_{i=1}^m\log \left(y_i^{\la-1}+y_i^{-\la-1}\right).
\end{equation}
Note that the third term in (\ref{like}) comes from the Jacobian of the transformation.
When $\la$ is given, maximizing (\ref{like}) with respect to $\bbe$ and $A$ coincides to maximizing the log-likelihood function of the classical Fay-Herriot model.
Hence, the value of profile likelihood function of $\la$ is easily computed, so that we may estimate $\la$ by grid search over a specified region or golden section method (Brent, 1973).    
Though the parameter space of $\la$ is $[0,\infty)$, it would be sufficient to consider the space $[0,\la_m]$ for moderately large $\la_m$.

For asymptotic properties of the estimator $\bphih$, we assume the following conditions.

\begin{as}\label{as:asymp}
\item[1.]
There exist $\underline{D}$ and $\overline{D}$ independent to $m$ such that $\underline{D}\leq D_i\leq \overline{D}$ for $i=1,\ldots,m$.

\item[2.]
$\max_{i=1,\ldots,m}\x_i^t(\sum_{j=1}^m\x_j\x_j^t)^{-1}\x_i=O(m^{-1})$.
\end{as}

These conditions are usually assumed in the context of small area estimation, see Datta and Lahiri (2000) and Butar and Lahiri (2003).
Under these conditions, we have the following lemma.

\begin{lem}\label{lem:asymp}
Under Assumption \ref{as:asymp}, as $m\to\infty$, $\sqrt{m}(\bphih-\bphi)$ asymptotically follows the multivariate normal distribution $N(\0,\V(\bphi))$ with a covariance matrix $\V(\bphi)$, and it holds $\E[\bphih-\bphi]=m^{-1}\b({\bphi})+o(m^{-1})$ with a smooth function $\b(\bphi)$. 
\end{lem}

The asymptotic normality of $\bphih$ immediately follows from Sugasawa and Kubokawa (2015). 
Moreover, from the proof of Theorem 1 in Lohr and Rao (2009), the bias $\b(\bphi)$ can be expressed by partial derivatives of $L(\bphi)$ given in (\ref{like}), so that the latter statement in Lemma \ref{lem:asymp} follows.

Other estimators of $A$ are the restricted maximum likelihood estimator (Jiang, 1996), the Prasad-Rao estimator (Prasad and Rao, 1990), the Fay-Herriot estimator (Fay and Herriot, 1979) and the adjusted maximum likelihood estimator (Li and Lahiri, 2010).
These methods can be easily implemented and their asymptotic properties are discussed in Sugasawa and Kubokawa (2015).
However, for simplicity, we do not treat these estimators in this paper.


\subsection{Mean squared error of the empirical best predictor}
In small area estimation, mean squared errors (MSEs) of small area estimators are used for risk evaluation, and their importance has been addressed in many papers including Lahiri and Rao (1995) and Datta et al. (2005).
Following this convention, we evaluate the MSE of the empirical best predictor $\muh_i$.
To begin with, we note that the MSE can be decomposed as
\begin{align*}
{\rm MSE}_i\equiv\E[(\muh_i-\mu_i)^2]
&=\E[(\mut_i-\mu_i)^2]+\E[(\muh_i-\mut_i)^2]\\
&\equiv g_{1i}(\bphi)+g_{2i}(\bphi),
\end{align*}
because $\mut_i=E[\mu_i|y_i]$ is the conditional expectation.
In what follows, we use the explicit notation $\mut_i(y_i,\bphi)$ instead of $\mut_i$ if necessary.
The first term $g_{1i}(\bphi)$ is expressed as
$$
g_{1i}(\bphi)=\E\left[\left\{\mut_i(\x_i^t\bbe+v_i+\ep_i,\bphi)-h_{\la}^{-1}(\x_i^t\bbe+v_i)\right\}^2\right],
$$
which has no analytical expression.
The direct Monte Carlo integration by generating random samples of $v_i$ and $\ep_i$ requires a large computational burden because we need another Monte Carlo integration for computing $\mut_i$ for each sample $(v_i,\ep_i)$.
However, as shown in the Appendix, it turns out to have the following more simple expression of $g_{1i}(\bphi)$: 
\begin{equation}\label{g1}
g_{1i}(\bphi)=\E\left[\{h_{\la}^{-1}(\x_i^t\bbe+z_1)\}^2-h_{\la}^{-1}(\x_i^t\bbe+c_{1i}z_1+c_{2i}z_2)h_{\la}^{-1}(\x_i^t\bbe+c_{1i}z_1-c_{2i}z_2)\right],
\end{equation}
where $z_1, z_2 \sim N(0,A)$, $c_{1i}=\sqrt{(1+a_i)/2}$, and $c_{2i}=\sqrt{(1-a_i)/2}$ for $a_i=A/(A+D_i)$.
Hence, $g_{1i}(\bphi)$ can be easily calculated by generating a large number of random samples of $z_1$ and $z_2$.
On the other hand, the second term $g_{2i}(\bphi)$ can be evaluated as the following lemma, where the proof is given in the Appendix.

\begin{lem}\label{lem:g2}
Under Assumption \ref{as:asymp}, it holds 
$$
g_{2i}(\bphi)
=\frac1m\tr\left\{\V(\bphi)\E\left[\frac{\partial\mut_i}{\partial\bphi}\frac{\partial\mut_i}{\partial\bphi^t}\right]\right\}+o(m^{-1}).
$$
\end{lem}

Since the MSE depends on unknown parameter $\bphi$, we need to estimate it for practical use.
To this end, we obtain a second-order unbiased estimator of the MSE.
Here, an estimator $\Bh$ is called second order unbiased if $\E[\Bh]=B+o(m^{-1})$.
From lemma \ref{lem:g2}, it follows that $g_{2i}(\bphi)=m^{-1}c_1(\bphi)+o(m^{-1})$ with the smooth function $c_1(\bphi)$, thereby the plug-in estimator $g_{2i}(\bphih)$ is second-order unbiased.
However, the plug-in estimator $g_{1i}(\bphih)$ has a second-order bias since $g_{1i}(\bphi)=O(1)$, so that we need to correct the bias.
Hence, we propose the parametric bootstrap method to correct the bias of $g_{1i}(\bphih)$ and computing $g_{2i}(\bphih)$.
The procedure is given in the following.

\ \\
{\bf Parametric bootstrap method for the MSE estimation}
\begin{itemize}
\item[1.] 
Generate bootstrap samples $y_i^{\ast }$ from the estimated model;
\begin{equation}\label{boot.samp}
h_{\lah}(y_i^{\ast })=\x_i^t\bbeh+v_i^{\ast}+\ep_i^{\ast }, \ \ \ i=1,\ldots,m,
\end{equation}
where $\ep_i^{\ast }$ and $v_i^{\ast }$ are generated from $N(0,D_i)$ and $N(0,\Ah)$, respectively

\item[2.]
Based on $(y_i^{\ast},\x_i), \ i=1,\ldots,m$, compute the maximum likelihood estimate $\bphih^{\ast }$ and the predicted values of $\muh_i=\mut_i(y_i,\bphih)$ and $\muh_i^{\ast}=\mut_i(y_i,\bphih^{\ast })$.

\item[3.]
Derive  the bootstrap estimates of $g_{1i}$ and $g_{2i}$ via
\begin{align*}
&g_{1i}^{\rm bc}(\bphih)=2g_{1i}(\bphih)-E^{\ast}\left[g_{1i}(\bphih^{\ast })\right], \ \ \ \ g_{2i}^{\ast}(\bphih)=E^{\ast}\left[\left(\muh_i^{\ast }-\muh_i\right)^2\right]
\end{align*}
where $\muh_i^{\ast }=h_{\lah}^{-1}(\x_i^t\bbeh+v_i^{\ast})$ and $E^{\ast}[\cdot]$ denotes the expectation with respect to the bootstrap samples generated from (\ref{boot.samp}).
The second-order unbiased estimator of the MSE based on the parametric bootstrap is given by
\begin{equation}\label{estMSE}
\widehat{{\rm MSE}_i}=g_{1i}^{\rm bc}(\bphih)+g_{2i}^{\ast}(\bphih).
\end{equation}
\end{itemize}

\ \\
The resulting MSE estimator (\ref{estMSE}) is second-order unbiased as shown in the following theorem, which is proved in the Appendix.

\begin{thm}\label{thm:mse}
Let $\widehat{{\rm MSE}_i}$ be the parametric bootstrap MSE estimator given in {\rm (\ref{estMSE})}.
Then, under Assumption \ref{as:asymp}, we have
$$
\E\left[\widehat{{\rm MSE}_i}\right]={\rm MSE}_i+o(m^{-1}),
$$
where the expectation is taken with respect to $y_i$'s following the model (\ref{TFH}).
\end{thm}

\medskip
In (\ref{estMSE}), the bias correction of $g_{1i}(\bphih)$ is carried out via using the additive form $g_{1i}^{\rm bc}(\bphih)=2g_{1i}(\bphih)-E^{\ast}[g_{1i}(\bphih^{\ast })]$, where $E^{\ast}$ denotes the expectation with respect to bootstrap samples.
Hall and Maiti (2006) suggested other bias-correcting methods including a multiplicative bias correcting method of the form $g_{1i}(\bphih)^2/E^{\ast}[g_{1i}(\bphih^{\ast })]$.
The multiplicative form for bias correction can avoid negative estimates of the MSE while the additive form for bias correction gives negative estimates of the MSE with a positive probability.
Although those bias corrections give second-order unbiased estimates of $g_{1i}(\bphi)$, in this paper, we use the additive-type bias correction, because it has been frequently used in the literatures (e.g. Butar and Lahiri, 2003).

\section{Simulation Studies}\label{sec:sim}


\subsection{Evaluation of prediction errors}\label{sec:comp}

We evaluated prediction errors of the proposed PTFH model and some existing models.
As a data generating process, we considered the following PTFH model:
\begin{equation}\label{dgp}
h_{\la}(y_i)=\beta_0+\beta_1x_i+v_i+\ep_i, \ \ \ i=1,\ldots,30,
\end{equation}
where $v_i\sim N(0,A)$, $\ep_i\sim N(0,D_i)$ with $\beta_0=1, \beta_1=1$ and $A=1.5$.
For $\la$, we treated the four cases $\la=0.1,0.4,0.7$ and $1.0$.
The covariates $x_i$ were initially generated from the uniform distribution on $(0,4)$ and fixed in simulation runs.
Concerning sampling variance $D_i$, we divided 30 areas into $5$ groups (from $G_1$ to $G_5$), and areas within the same group have the same $D_i$ value.
The $D_i$-pattern we considered was $(0.2,0.4,0.6,0.8,1.0)$.
The true small area parameters are $\mu_i=h_{\la}^{-1}(\beta_0+\beta_1x_i+v_i)$.

For comparison, we considered the log-transformed FH (log-FH) model and the traditional Fay-Herriot (FH) model, which are described as 
\begin{align*}
\text{log-FH:} \ \ \ 
&\log y_i=\beta_0+\beta_1x_i+v_i+\ep_i\\
\text{FH:} \ \ \ 
&y_i=\beta_0+\beta_1x_i+v_i+\ep_i.
\end{align*}
It is noted that the data generating process (\ref{dgp}) get close to log-FH as $\la$ gets smaller.

Since we do not known the true $D_i$ in practice, we computed the estimates of $\mu_i$ with estimated $D_i$ as investigated in Bell (2008).
To this end, we generated the auxiliary observation $z_{ik}$ from the model:
\begin{equation}\label{aux}
h_{\la}(z_{ik})=\ep_{ik}, \ \ \ i=1,\ldots,30, \ \ k=1,\ldots,10,
\end{equation}
where $\ep_{ij}\sim N(0,D_i)$.
In applying log-FH and FH, we computed the estimates of $D_i$ as the sampling variances of $\{\log z_{i1},\ldots,\log z_{i10}\}$ and $\{z_{i1},\ldots,z_{i10}\}$, respectively.
Then we computed the estimates of $\mu_i$ using EBLUP in FH and the bias-corrected estimator used in Slud and Maiti (2006) in log-FH, where the model parameters $\be_0, \be_1$ and $A$ are estimated via the maximum likelihood method.
For fitting PTFH, we first define
\begin{equation*}
D_i\equiv D_i(\la)=\frac{1}{9}\sum_{k=1}^{10}\left\{h_\la(z_{ik})-\overline{h_{\la}(z)}_i\right\}^2, \ \ \ \  \overline{h_{\la}(z)}_i=\frac1{10}\sum_{k=1}^{10}h_{\la}(z_{ik}), 
\end{equation*}
so that we regard $D_i$ as a function of $\la$ and replace $D_i$ with $D_i(\la)$ in (\ref{like}).
Since $D_i(\la)$ can be immediately computed under the given $\la$, we can maximize the profile likelihood function of $\la$ in the similar manner to that presented in Section \ref{sec:est}.
Once the estimate $\lah$ is computed, $D_i$ can be calculated as $D_i(\lah)$.

Based on $R=10000$ simulation runs, we computed the coefficient of variation (CV) and the absolute relative bias (ARB), defined as 
$$
\text{CV}_i=\sqrt{\frac1R\sum_{r=1}^{R}\frac{\big(\muh_i^{(r)}-\mu_i^{(r)}\big)^2}{\mu_i^{(r)2}}} \ \ \ \ \ \text{and} \ \ \ \ 
\text{ARB}_i=\bigg|\frac1R\sum_{r=1}^{R}\frac{\muh_i^{(r)}-\mu_i^{(r)}}{\mu_i^{(r)}}\bigg|,
$$
where $\mu_i^{(r)}$ is the true value and $\muh^{(r)}$ is the estimated value from PTFH, log-FH or FH, in the $r$th iteration.
Table \ref{tab:comp} shows the percent CV and ARB averaged within the same groups for each case of $\la$.
For comparison, we also show the results for PTFH with true $D_i$ values, denoted by PTFH-t in Table \ref{tab:comp}.

From Table \ref{tab:comp}, we can observe that difference CV or ARB between PTFH-t and PTFH tends to be large when $\la$ is small and $D_i$ is large while tow methods perform similarly when $\la$ is large or $D_i$ is small.
Moreover, it is revealed that the larger $D_i$ would lead to larger CV and ARB values in all the methods.
Concerning comparison among PTFH, log-FH and FH, it can be seen that PTFH performs better than FH except for $\la=0.9$, and PTFH performs better than log-FH except for $\la=0.1$.
Moreover, the differences between PTFH and log-FH in $\la=0.1$ get larger as $D_i$ gets larger. 
Regarding PTFH-t, it performs best in most cases.
However, it is observed that log-FH and FH produce more accurate estimates than PTFH-t in some cases.

We next investigated the prediction errors when the true distribution of $v_i$ is not normal. 
Here, we considered a $t$-distribution with $5$ degrees of freedom for $v_i$, where the variance is scaled to $A$, and the other settings for the data generation are the same as (\ref{dgp}).
Under the scenario, we again computed the values of CV and ARB of the four methods based on $10000$ simulation runs, and the results are reported in Table \ref{tab:comp2}.
It is observed that the simulated RMSE values in Table \ref{tab:comp2} are larger than those in Table \ref{tab:comp} due to misspecification of the distribution of $v_i$.
However, relationships of CV and ARB among three methods are similar to Table \ref{tab:comp}.

\begin{table}[!htb]
\caption{Simulated percent coefficient of variation (CV) and absolute relative biases (ARB) of the parametric transformed Fay-Herriot with use of true $D_i$ (PTFH-t) and estimated $D_i$ (PTFH), the log-transformed Fay-Herriot (log-FH) model, and the Fay-Herriot (FH) model under $\la=0.1,0.4,0.7$ and $1.0$.
\label{tab:comp}
}
\vspace{0cm}
\begin{center}
$
{\renewcommand\arraystretch{1.1}\small
\begin{array}{c@{\hspace{2mm}}c@{\hspace{2mm}}
c@{\hspace{2mm}}c@{\hspace{2mm}}c@{\hspace{2mm}}
c@{\hspace{2mm}}c@{\hspace{2mm}}c@{\hspace{2mm}}
c@{\hspace{2mm}}c@{\hspace{2mm}}c@{\hspace{2mm}}
c@{\hspace{2mm}}c@{\hspace{2mm}}c@{\hspace{2mm}}
}
\hline
&& \multicolumn{4}{c}{\text{CV}} && \multicolumn{4}{c}{\text{ARB}}\\
 & \text{method} & 0.1 &0.4 & 0.7 & 1.0 && 0.1 & 0.4 & 0.7 & 1.0 \\
 \hline
\text{$G_1$}  & \text{PTFH-t} & 46.49 & 33.28 & 23.49 & 17.95 &  & 13.67 & 6.89 & 3.69 & 2.43 \\
 & \text{PTFH} & 47.46 & 32.90 & 23.12 & 17.53 &  & 13.97 & 6.89 & 3.51 & 2.21 \\
 & \text{log-FH} & 47.11 & 35.34 & 28.63 & 23.57 &  & 13.16 & 4.18 & 3.85 & 2.39 \\
 & \text{FH} & 50.24 & 32.94 & 22.84 & 16.87 &  & 9.74 & 3.82 & 1.78 & 1.13 \\
 \hline
\text{$G_2$}  & \text{PTFH-t} & 63.92 & 47.64 & 37.09 & 31.61 &  & 20.20 & 11.85 & 8.13 & 6.69 \\
 & \text{PTFH} & 66.32 & 47.84 & 36.92 & 31.38 &  & 21.28 & 12.24 & 8.31 & 6.74 \\
 & \text{log-FH} & 66.79 & 53.25 & 46.70 & 41.76 &  & 22.25 & 15.14 & 13.96 & 13.86 \\
 & \text{FH} & 82.25 & 52.67 & 38.61 & 31.11 &  & 19.95 & 9.16 & 6.27 & 5.98 \\
 \hline
\text{$G_3$}  & \text{PTFH-t} & 75.92 & 59.42 & 48.51 & 40.60 &  & 25.67 & 16.79 & 12.85 & 9.59 \\
 & \text{PTFH} & 79.86 & 60.35 & 48.22 & 40.35 &  & 27.67 & 17.74 & 13.15 & 9.67 \\
 & \text{log-FH} & 77.70 & 60.99 & 53.25 & 47.15 &  & 26.56 & 15.97 & 13.04 & 11.30 \\
 & \text{FH} & 116.09 & 74.04 & 53.70 & 40.88 &  & 30.55 & 17.10 & 11.77 & 9.25 \\
 \hline
\text{$G_4$}  & \text{PTFH-t} & 86.92 & 67.82 & 53.66 & 44.73 &  & 32.29 & 20.84 & 14.11 & 10.82 \\
 & \text{PTFH} & 92.97 & 69.13 & 53.38 & 43.83 &  & 35.12 & 22.13 & 14.43 & 10.63 \\
 & \text{log-FH} & 86.81 & 65.36 & 53.46 & 46.36 &  & 30.88 & 14.42 & 8.29 & 8.94 \\
 & \text{FH} & 156.12 & 91.68 & 61.04 & 45.61 &  & 45.29 & 25.10 & 15.15 & 11.25 \\
 \hline
\text{$G_5$}  & \text{PTFH-t} & 92.90 & 72.74 & 59.86 & 49.86 &  & 33.87 & 22.99 & 17.30 & 13.04 \\
 & \text{PTFH} & 101.81 & 75.26 & 60.39 & 49.54 &  & 37.62 & 24.73 & 18.07 & 13.24 \\
 & \text{log-FH} & 95.91 & 71.73 & 61.69 & 53.58 &  & 34.91 & 20.57 & 15.30 & 12.87 \\
 & \text{FH} & 198.30 & 112.23 & 73.57 & 53.02 &  & 57.58 & 31.04 & 19.05 & 13.93 \\

 \hline
 \end{array}
}
$
\end{center}
\end{table}

\begin{table}[!htb]
\caption{Simulated percentage coefficient of variation (CV) and percentage absolute relative biases (ARB) of the parametric transformed Fay-Herriot with use of true $D_i$ (PTFH-t) and estimated $D_i$ (PTFH), the log-transformed Fay-Herriot (log-FH) model, and the Fay-Herriot (FH) model under $\la=0.1,0.4,0.7$ and $1.0$, when the distribution of $v_i$ is a $t$-distribution with 5 degrees of freedom.
\label{tab:comp2}
}
\vspace{0cm}
\begin{center}
$
{\renewcommand\arraystretch{1.1}\small
\begin{array}{c@{\hspace{2mm}}c@{\hspace{2mm}}
c@{\hspace{2mm}}c@{\hspace{2mm}}c@{\hspace{2mm}}
c@{\hspace{2mm}}c@{\hspace{2mm}}c@{\hspace{2mm}}
c@{\hspace{2mm}}c@{\hspace{2mm}}c@{\hspace{2mm}}
c@{\hspace{2mm}}c@{\hspace{2mm}}c@{\hspace{2mm}}
}
\hline
&& \multicolumn{4}{c}{\text{CV}} && \multicolumn{4}{c}{\text{ARB}}\\
 & \text{method} & 0.1 &0.4 & 0.7 & 1.0 && 0.1 & 0.4 & 0.7 & 1.0 \\
 \hline
\text{$G_1$}  & \text{PTFH-t} & 47.19 & 39.58 & 34.16 & 28.94 &  & 14.10 & 9.77 & 7.31 & 5.43 \\
 & \text{PTFH} & 48.42 & 39.84 & 34.40 & 28.34 &  & 14.70 & 10.07 & 7.38 & 5.18 \\
 & \text{log-FH} & 48.09 & 41.08 & 39.98 & 35.31 &  & 14.00 & 8.14 & 4.95 & 5.65 \\
 & \text{FH} & 51.54 & 41.09 & 33.00 & 27.84 &  & 10.26 & 6.92 & 5.18 & 4.28 \\
 \hline
\text{$G_2$}  & \text{PTFH-t} & 66.06 & 53.12 & 41.12 & 35.83 &  & 21.41 & 14.08 & 8.99 & 6.92 \\
 & \text{PTFH} & 68.38 & 53.52 & 40.86 & 35.18 &  & 22.59 & 14.78 & 9.10 & 6.80 \\
 & \text{log-FH} & 67.79 & 56.17 & 46.04 & 43.20 &  & 21.84 & 13.32 & 7.32 & 6.73 \\
 & \text{FH} & 83.45 & 58.51 & 41.73 & 34.32 &  & 21.03 & 12.63 & 7.67 & 6.18 \\
 \hline
\text{$G_3$}  & \text{PTFH-t} & 79.92 & 62.21 & 49.91 & 42.61 &  & 25.50 & 16.81 & 10.91 & 8.69 \\
 & \text{PTFH} & 83.75 & 62.26 & 47.41 & 41.62 &  & 27.39 & 17.60 & 11.05 & 8.66 \\
 & \text{log-FH} & 83.53 & 65.45 & 55.47 & 52.08 &  & 27.40 & 18.20 & 12.93 & 13.22 \\
 & \text{FH} & 121.91 & 73.92 & 50.11 & 41.05 &  & 31.49 & 17.11 & 10.35 & 8.25 \\
 \hline
\text{$G_4$}  & \text{PTFH-t} & 89.82 & 78.86 & 61.81 & 53.21 &  & 30.06 & 21.02 & 15.63 & 12.03 \\
 & \text{PTFH} & 98.67 & 75.36 & 59.27 & 51.18 &  & 32.81 & 22.26 & 15.83 & 12.01 \\
 & \text{log-FH} & 99.13 & 81.96 & 62.46 & 57.54 &  & 31.16 & 19.98 & 13.48 & 11.01 \\
 & \text{FH} & 155.72 & 108.51 & 65.22 & 52.10 &  & 44.14 & 25.07 & 15.96 & 12.80 \\
 \hline
\text{$G_5$}  & \text{PTFH-t} & 104.87 & 86.53 & 77.05 & 71.23 &  & 34.72 & 26.73 & 22.00 & 17.04 \\
 & \text{PTFH} & 123.97 & 86.69 & 74.22 & 64.63 &  & 38.87 & 28.84 & 22.83 & 17.06 \\
 & \text{log-FH} & 120.17 & 87.54 & 75.41 & 67.96 &  & 34.96 & 22.70 & 15.61 & 11.36 \\
 & \text{FH} & 224.09 & 128.60 & 87.85 & 69.25 &  & 61.85 & 40.85 & 29.11 & 21.47 \\
 \hline
 \end{array}
}
$
\end{center}
\end{table}


\subsection{Finite sample performance of the MSE estimator}
We next investigated a finite sample performance of the MSE estimator (\ref{estMSE}).
Following Datta, Rao and Smith (2005), we considered the following data generating process without covariates:
$$
h_{\la}(y_i)=\mu+v_i+\ep_i, \ \ \ \ i=1,\ldots,30,
$$
with $\mu=0$, $v_i\sim N(0,A)$ with $A=1$ and $\ep_i\sim N(0,D_i)$.
As a value of $\la$, we considered the three cases $\la=0.2, 0.6, 1.0$.
For setting of $D_i$, we divided $D_i$'s into five groups $G_1,\ldots,G_5$, where $D_i$'s were the same values over the same group, and the following three patterns of $D_i$'s were considered:
$$
(a)\ 0.3, 0.4, 0.5, 0.6, 0.7, \ \ (b) \ 0.2, 0.4, 0.5, 0.6, 2.0, \ \ (c)\ 0.1, 0.4, 0.5, 0.6, 4.0.
$$
Based on $R_1=5000$ simulation runs, we calculated the simulated values of the MSE as
$$
{\rm MSE}_i=\frac1{R_1}\sum_{r=1}^{R_1}(\muh_i^{(r)}-\mu_i^{(r)})^2, \ \ \ \ \ \mu_i^{(r)}=h_{\la}^{-1}(\beta_0+v_i^{(r)})
$$
where $\muh_i^{(r)}$ and $v_i^{(r)}$ are the predicted value and the realized value of $v_i$ in the $r$-th iteration.
Then based on $R_2=2000$ simulation runs, we calculated the relative bias (RB) and the coefficient of variation (CV) defined as 
\begin{align*}
&{\rm RB}_i=\frac1{R_2}\sum_{r=1}^{R_2}\left(\widehat{{\rm MSE}_i}^{(r)}-{\rm MSE}_i\right)/{\rm MSE}_i, \\
&{\rm CV}_i^2=\frac1{R_2}\sum_{r=1}^{R_2}\left(\widehat{{\rm MSE}_i}^{(r)}-{\rm MSE}_i\right)^2/{\rm MSE}_i^2.
\end{align*}
For calculation of the MSE estimates in each iteration, we used 100 bootstrap replication for the MSE estimator and 10000 Monte Carlo samples for computing $g_{1i}$.
We also investigated the performance of the MSE estimator when we used the estimated sampling variances instead of known $D_i$.
To this end, similarly to the previous section, we generated the auxiliary observation from (\ref{aux}), and calculate $D_i$'s using these data in each simulation run.
Based on the same number of simulation runs, we computed the values of RB and CV.
Table \ref{tab:RB} and Table \ref{tab:CV} show the maximum, mean and minimum values of RB and CV within the same group.
In both tables, the simulated values of RB and CV of the MSE estimator with estimated $D_i$ are given in the parenthesis.
It is seen that the proposed MSE estimator with known $D_i$ provides reasonable estimated values in almost all cases in terms of both RB and CV.
On the other hand, the MSE estimator with estimated $D_i$ performs worse than the MSE estimator with known $D_i$ since the former estimator is affected by the variability of estimating $D_i$.
Moreover, it is observed that performances of both MSE estimators get better in the order of Pattern (a), (b) and (c).

\begin{table}[!htb]
\caption{The percentage RB values of the MSE estimator with known $D_i$ and unknown $D_i$ (Parenthesis) in each group.
\label{tab:RB}
}
\vspace{-0.4cm}
\begin{center}
$
{\renewcommand\arraystretch{1.1}\small
\begin{array}{c@{\hspace{3mm}}c@{\hspace{4mm}}
c@{\hspace{3mm}}c@{\hspace{3mm}}c@{\hspace{3mm}}
c@{\hspace{2mm}}c@{\hspace{2mm}}c@{\hspace{2mm}}
c@{\hspace{2mm}}c@{\hspace{2mm}}c@{\hspace{2mm}}
c@{\hspace{2mm}}c@{\hspace{2mm}}c@{\hspace{2mm}}
}
\hline
&&\multicolumn{3}{c}{{\rm Pattern\ (a)}} &  \multicolumn{3}{c}{{\rm Pattern\ (b)}} &  \multicolumn{3}{c}{{\rm Pattern\ (c)}} \\ 
 & \la & 0.2 & 0.6 & 1.0 & 0.2 & 0.6 & 1.0 & 0.2 & 0.6 & 1.0 \\
 \hline
\text{max} & G_1 & 6.3 & 17.9 & 19.5 & 3.9 & 7.3 & 9.0 & 7.2 & 1.3 & 2.7 \\
 &  & (28.1 )&( 44.3 )&( 51.4 )&( 98.4 )&( 80.8 )&( 82.6 )&( 113.5 )&( 47.1 )&( 44.1) \\
 & G_2 & 4.9 & 15.1 & 17.2 & 4.4 & 4.4 & 5.3 & 4.1 & -3.6 & -4.9 \\
 &  & (44.6 )&( 69.2 )&( 78.8 )&( 46.4 )&( 107.2 )&( 99.8 )&( 37.3 )&( 29.8 )&( 20.8) \\
 & G_3 & 12.5 & 10.4 & 14.6 & -2.8 & 3.3 & 4.6 & -3.4 & -4.1 & -5.0 \\
 &  & (32.9 )&( 40.8 )&( 47.0 )&( 24.2 )&( 12.5 )&( 10.6 )&( 11.3 )&( 37.5 )&( 29.7 )\\
 & G_4 & 6.4 & 12.6 & 17.6 & -2.6 & 6.7 & 8.9 & -5.0 & -4.5 & -4.8 \\
 &  & (33.8 )&( 28.3 )&( 40.2 )&( 12.1 )&( 8.8 )&( 8.0 )&( 16.1 )&( 84.4 )&( 75.7) \\
 & G_5 & 8.7 & 12.4 & 16.3 & -1.5 & 2.8 & 6.1 & -5.3 & -4.2 & -1.6 \\
 &  & (15.9 )&( 39.1 )&( 52.1 )&( 6.2 )&( 43.9 )&( 43.6 )&( 4.6 )&( 26.0 )&( 19.5 )\\
 \hline
\text{mean} & G_1 & 2.5 & 7.4 & 8.6 & -1.9 & 1.4 & 2.6 & -2.7 & -3.3 & -3.1 \\
 &  &( 22.4 )&( 31.4 )&( 36.9 )&( 33.5 )&( 32.3 )&( 33.9 )&( 47.6 )&( 27.7 )&( 27.0 )\\
 & G_2 & -1.0 & 7.8 & 9.8 & -3.4 & -2.0 & -2.0 & -6.0 & -5.9 & -7.3 \\
 &  &( 21.9 )&( 33.4 )&( 39.9 )&( 15.5 )&( 21.5 )&( 18.8 )&( 20.2 )&( 4.8 )&( -0.6 )\\
 & G_3 & 5.2 & 1.9 & 5.2 & -6.3 & -1.7 & -0.4 & -6.1 & -8.4 & -9.1 \\
 &  &( 17.7 )&( 21.1 )&( 28.7 )&( 11.2 )&( 3.6 )&( 2.4 )&( -0.1 )&( 10.4 )&( 4.6) \\
 & G_4 & 3.3 & 6.4 & 9.9 & -8.0 & -0.4 & 1.5 & -7.4 & -7.5 & -7.9 \\
 &  &( 20.4 )&( 17.7 )&( 26.9 )&( 2.2 )&( -2.8 )&( -2.2 )&( -5.3 )&( 15.6 )&( 10.9) \\
 & G_5 & 1.4 & 5.7 & 9.9 & -6.0 & -0.2 & 2.4 & -7.0 & -6.8 & -6.4 \\
 &  &( 11.1 )&( 22.1 )&( 32.8 )&( -3.0 )&( 6.4 )&( 7.2 )&( -6.8 )&( -1.2 )&( -4.4) \\
 \hline
\text{min} & G_1 & -4.9 & -3.2 & -2.4 & -6.9 & -5.2 & -4.5 & -8.5 & -6.9 & -6.6 \\
 &  &( 17.2 )&( 15.0 )&( 20.2 )&( 13.7 )&( 16.1 )&( 17.3 )&( 21.5 )&( 17.1 )&( 17.0) \\
 & G_2 & -5.8 & -0.9 & 2.8 & -7.5 & -7.0 & -7.5 & -9.6 & -7.8 & -9.9 \\
 &  &( 11.5 )&( 14.2 )&( 18.6 )&( 2.7 )&( -2.7 )&( -4.7 )&( 8.1 )&( -5.2 )&( -8.8 )\\
 & G_3 & 0.9 & -8.2 & -6.7 & -8.7 & -6.6 & -6.5 & -10.3 & -10.4 & -11.3 \\
 &  &( 7.0 )&( 13.2 )&( 19.4 )&( -5.0 )&( -4.4 )&( -4.7 )&( -7.3 )&( -9.4 )&( -13.2) \\
 & G_4 & -1.9 & -3.0 & -0.2 & -11.9 & -5.8 & -3.9 & -9.1 & -11.2 & -11.9 \\
 &  &( 10.0 )&( 6.9 )&( 14.1 )&( -7.0 )&( -11.6 )&( -9.5 )&( -12.6 )&( -16.1 )&( -18.2) \\
 & G_5 & -3.4 & -6.4 & -2.4 & -7.9 & -5.4 & -2.8 & -8.5 & -9.3 & -9.8 \\
 &  &( 6.4 )&( -0.6 )&( 8.8 )&( -7.3 )&( -10.4 )&( -9.7 )&( -16.1 )&( -18.4 )&( -20.0) \\
\hline
\end{array}
}
$
\end{center}
\end{table}

\begin{table}[!htb]
\caption{The CV values of the MSE estimator with known $D_i$ and unknown $D_i$ (Parenthesis) in each group.
\label{tab:CV}
}
\vspace{-0.2cm}
\begin{center}
$
{\renewcommand\arraystretch{1.1}\small
\begin{array}{c@{\hspace{3mm}}c@{\hspace{4mm}}
c@{\hspace{3mm}}c@{\hspace{3mm}}c@{\hspace{3mm}}
c@{\hspace{2mm}}c@{\hspace{2mm}}c@{\hspace{2mm}}
c@{\hspace{2mm}}c@{\hspace{2mm}}c@{\hspace{2mm}}
c@{\hspace{2mm}}c@{\hspace{2mm}}c@{\hspace{2mm}}
}
\hline
&&\multicolumn{3}{c}{{\rm Pattern\ (a)}} &  \multicolumn{3}{c}{{\rm Pattern\ (b)}} &  \multicolumn{3}{c}{{\rm Pattern\ (c)}} \\ 
 & \la & 0.2 & 0.6 & 1.0 & 0.2 & 0.6 & 1.0 & 0.2 & 0.6 & 1.0 \\
 \hline
\text{max} & G_1 & 1.15 & 1.03 & 1.14 & 0.78 & 0.60 & 0.66 & 0.77 & 0.62 & 0.67 \\
 &  & (2.72) & (1.65) & (1.96) & (2.13) & (1.78) & (1.89) & (2.01) & (1.19) & (1.29) \\
 & G_2 & 1.17 & 1.33 & 1.50 & 0.55 & 0.73 & 0.71 & 0.54 & 0.50 & 0.54 \\
 &  & (2.25) & (2.50) & (3.42) & (1.20) & (2.44) & (2.49) & (0.91) & (0.84) & (0.82) \\
 & G_3 & 2.25 & 0.98 & 1.16 & 0.59 & 0.87 & 1.00 & 0.52 & 0.46 & 0.53 \\
 &  & (1.34) & (1.80) & (2.16) & (0.88) & (1.02) & (1.1) & (0.68) & (0.98) & (0.97) \\
 & G_4 & 1.02 & 1.26 & 1.54 & 0.46 & 0.84 & 0.98 & 0.51 & 0.50 & 0.57 \\
 &  & (1.44) & (1.40) & (2.08) & (0.70) & (0.94) & (1.06) & (0.66) & (1.72) & (1.70) \\
 & G_5 & 0.72 & 1.01 & 1.23 & 0.67 & 0.73 & 0.88 & 0.47 & 0.66 & 0.77 \\
 &  & (1.10) & (1.59) & (2.21) & (0.91) & (1.60) & (1.87) & (0.56) & (0.90) & (0.91) \\
 \hline
\text{mean} & G_1 & 0.99 & 0.74 & 0.82 & 0.51 & 0.50 & 0.55 & 0.45 & 0.41 & 0.43 \\
 &  & (1.05) & (1.33) & (1.57) & (1.07) & (1.07) & (1.15) & (1.10) & (0.86) & (0.90) \\
 & G_2 & 0.78 & 0.91 & 1.06 & 0.46 & 0.56 & 0.63 & 0.40 & 0.42 & 0.47 \\
 &  & (1.23) & (1.46) & (1.83) & (0.80) & (1.05) & (1.11) & (0.73) & (0.66) & (0.68) \\
 & G_3 & 1.02 & 0.86 & 1.03 & 0.46 & 0.61 & 0.71 & 0.42 & 0.41 & 0.45 \\
 &  & (1.08) & (1.27) & (1.60) & (0.76) & (0.81) & (0.88) & (0.60) & (0.70) & (0.72) \\
 & G_4 & 0.75 & 0.93 & 1.13 & 0.43 & 0.66 & 0.78 & 0.41 & 0.44 & 0.50 \\
 &  & (1.10) & (1.24) & (1.65) & (0.64) & (0.80) & (0.89) & (0.56) & (0.85) & (0.87) \\
 & G_5 & 0.71 & 0.92 & 1.12 & 0.52 & 0.66 & 0.77 & 0.41 & 0.49 & 0.56 \\
 &  & (0.99) & (1.36) & (1.83) & (0.70) & (0.94) & (1.05) & (0.53) & (0.69) & (0.72) \\
 \hline
\text{min} & G_1 & 0.61 & 0.58 & 0.65 & 0.37 & 0.41 & 0.45 & 0.35 & 0.35 & 0.36 \\
 &  & (0.95) & (0.93) & (1.12) & (0.69) & (0.86) & (0.91) & (0.71) & (0.73) & (0.76) \\
 & G_2 & 0.57 & 0.73 & 0.86 & 0.39 & 0.49 & 0.56 & 0.34 & 0.39 & 0.41 \\
 &  & (0.89) & (1.10) & (1.28) & (0.64) & (0.75) & (0.80) & (0.57) & (0.58) & (0.59) \\
 & G_3 & 0.63 & 0.79 & 0.88 & 0.40 & 0.52 & 0.60 & 0.37 & 0.39 & 0.43 \\
 &  & (0.95) & (1.11) & (1.35) & (0.61) & (0.72) & (0.80) & (0.54) & (0.55) & (0.58) \\
 & G_4 & 0.66 & 0.81 & 0.99 & 0.41 & 0.59 & 0.69 & 0.37 & 0.42 & 0.45 \\
 &  & (0.97) & (1.07) & (1.33) & (0.61) & (0.69) & (0.76) & (0.49) & (0.55) & (0.59) \\
 & G_5 & 0.68 & 0.80 & 1.00 & 0.43 & 0.58 & 0.69 & 0.38 & 0.42 & 0.47 \\
 &  & (0.92) & (1.06) & (1.40) & (0.59) & (0.74) & (0.83) & (0.50) & (0.55) & (0.59) \\
\hline
\end{array}
}
$
\end{center}
\end{table}

\section{Application to Survey Data in Japan}\label{sec:emp}
We consider an application of the proposed method together with some existing methods to the data from the Survey of Family Income and Expenditure (SFIE) in Japan. 
Especially, we used the data on the spending item `Health' and `Education' in the survey in 2014.
For the spending item `Health' and `Education', the annual average spending data at each capital city of 47 prefectures are available. 
The estimates are both unreliable since the sample sizes are around 50 for most prefectures.
As a covariate, we used data from the National Survey of Family Income and Expenditure (NSFIE) for 47 prefectures.
Since NSFIE is based on much larger sample than SFIE, the reported values are more reliable, but this survey has been implemented every five years.
Although the joint bivariate modeling of the two items `Health' and `Education' would be preferable as proposed in Benavent et al. (2016), we here consider applying univariate models separately to each item for simplicity.
In what follows, $y_i$ and $x_i$ denote the direct estimate (scaled by 1000) from SFIE and the covariate (reliable estimate) from NSFIE, respectively, on the item `Health' or `Education'.
For each survey data, we applied the PTFH model:
$$
h_{\la}(y_i)=\beta_{0}+\beta_{1}\log x_{i}+v_{i}+\ep_{i}, \ \ \ i=1,\ldots,47,
$$
where $v_i\sim N(0,A)$, $\ep_i\sim N(0,D_i)$ and $\be_0,\be_1$, $A$ and $\la$ are model parameters.
For comparison, we also applied the log-FH model corresponding $\la=0$ in the above model, and the classical Fay-Herriot model. 
The model parameters were estimated by the maximum likelihood method in all models.
For computing $D_i$ in each model, we used the past data for consecutive eight years from 2006 to 2013, which are denoted by $z_{it}$ for $t=1,\ldots,8$.
In the FH and log-FH models, we simply calculated the sampling variance of $\{z_{i1},\ldots,z_{i8}\}$ and $\{\log z_{i1},\ldots,\log z_{i8}\}$, respectively.
In the PTFH model, similarly to Section \ref{sec:comp}, we first maximize (\ref{like}) with $D_i=D_i(\la)$ and let $D_i=D_i(\lah)$.

\begin{figure}
\begin{center}
\includegraphics[width=6cm,clip]{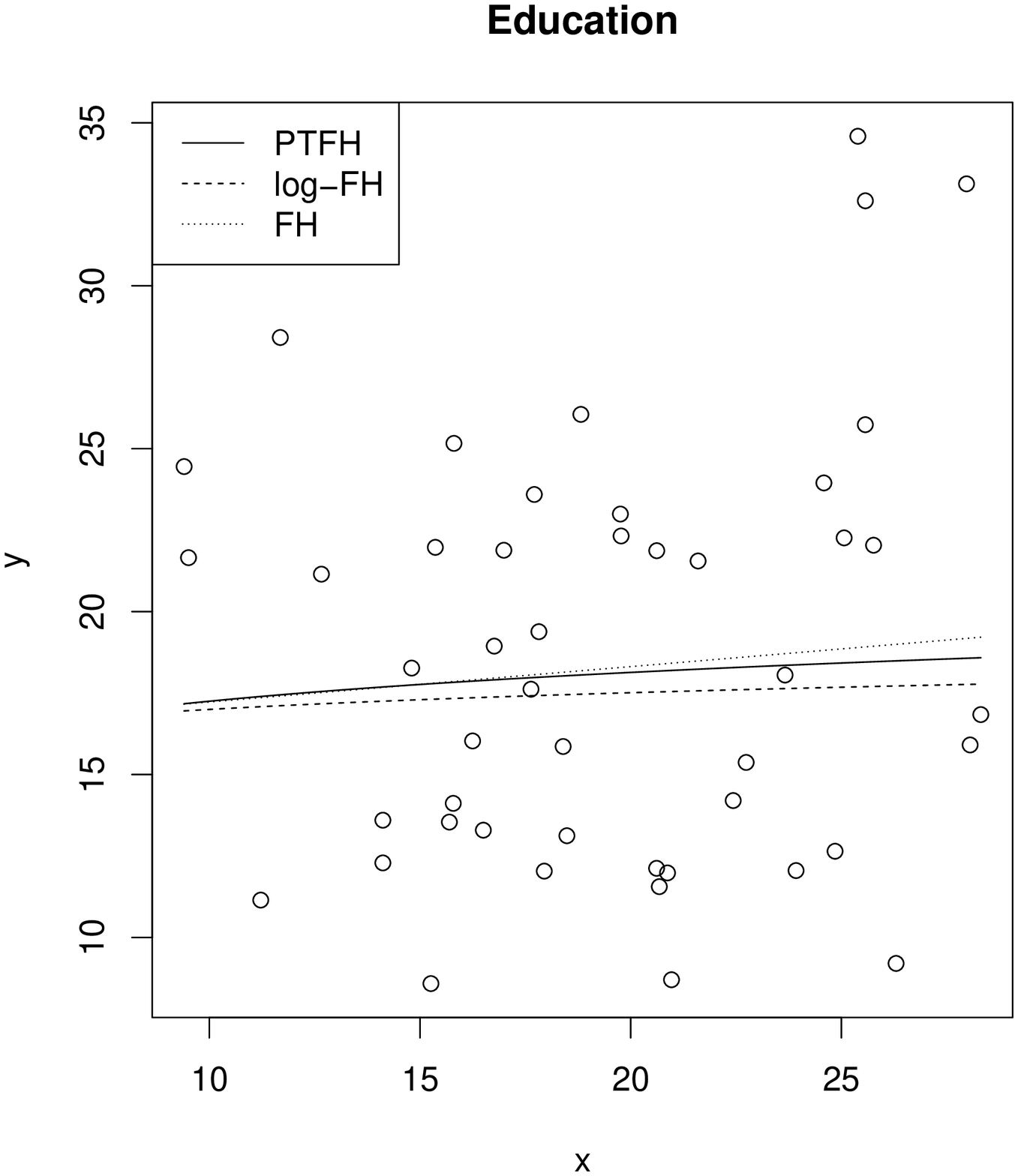}
\includegraphics[width=6cm,clip]{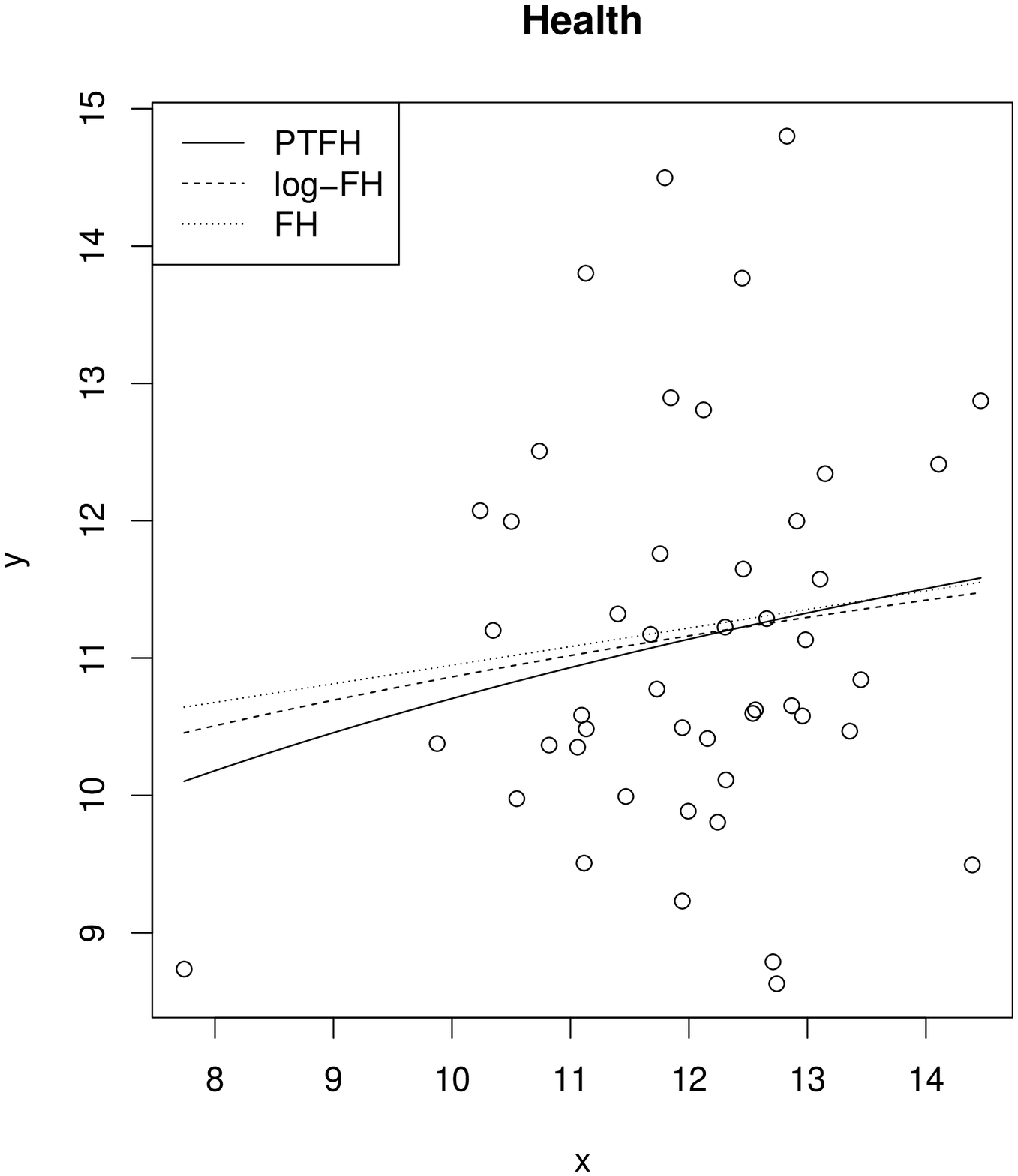}
\caption{The scatter plots of $(x_i,y_i)$ with estimated regression lines in the parametric transformed Fay-Herriot (PTFH) model, the log-transformed Fay-Herriot (log-FH) model and the classical Fay-Herriot (FH) model. 
\label{fig:app-reg}
}
\end{center}
\end{figure}

\begin{figure}
\begin{center}
\includegraphics[width=6cm,clip]{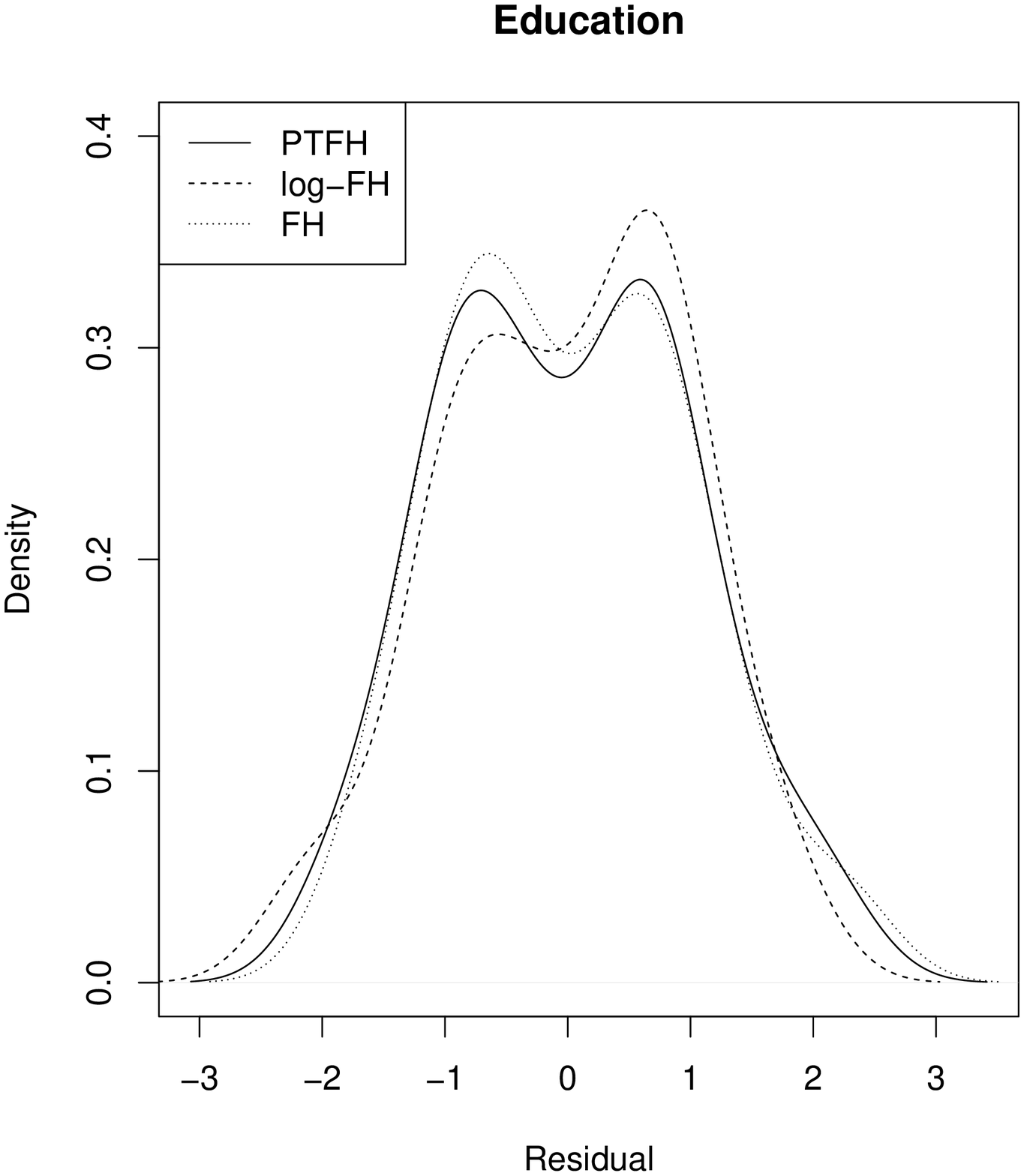}
\includegraphics[width=6cm,clip]{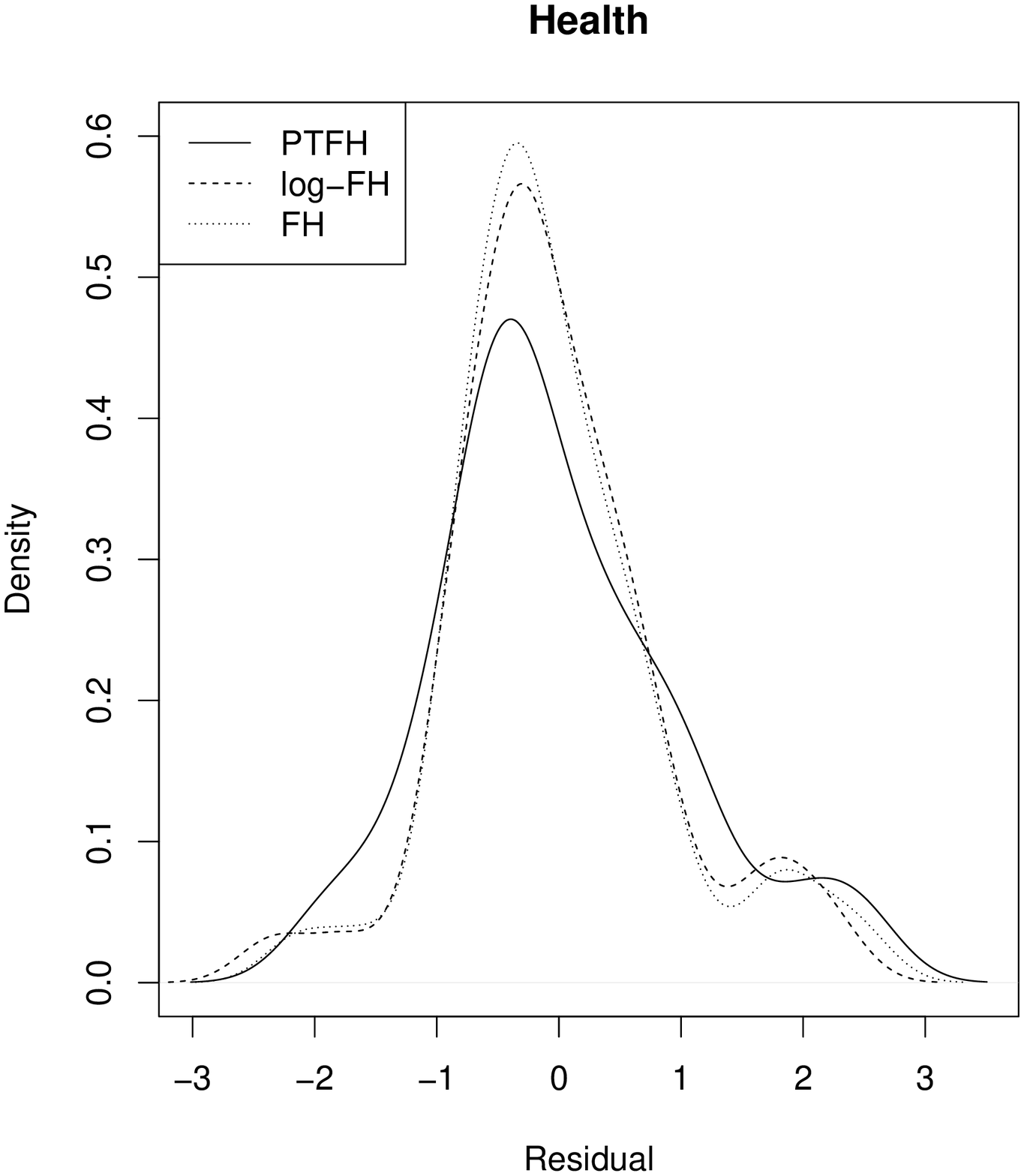}
\caption{The estimated density of standardized residuals in the parametric transformed Fay-Herriot (PTFH) model, the log-transformed Fay-Herriot (log-FH) model and the classical Fay-Herriot (FH) model. 
\label{fig:app-resid}
}
\end{center}
\end{figure}

In the PTFH  model, we have $\lah=0.59$ in ``Education" and $\lah=0.86$ in ``Health".
Moreover, based on based on $1000$ parametric bootstrap samples, we obtained $95\%$ confidence intervals of $\la$, $(0.20,1.16)$ in ``Education" and $(0.18,1.99)$ in ``Health", which indicate the log-transformation might not be appropriate.
In Figure \ref{fig:app-reg}, we present the estimated regression lines of the three models, noting that $y=h_{\lah}^{-1}(\beh_0+\beh_1x)$ in PTFH, and $y=\exp(\beh_0+\beh_1x)$ in log-FH.
From the figure, it is observed that all the regression lines are similar.
For assessing the suitability of normality assumptions of error terms, we computed the standardized residuals: $e_i=r_i/\sqrt{\hat{A}+D_i}$, where $r_i$ is the estimates of $v_i+\ep_i$, so that $r_i=h_{\lah}(y_i)-\beh_0-\beh_1x_i$ in PTFH, $r_i=\log y_i-\beh_0-\beh_1x_i$ in log-FH and $r_i=y_i-\beh_0-\beh_1x_i$ in FH, noting that $e_i$ asymptotically follows the standard normal distribution if the model specification is correct.
In Figure \ref{fig:app-resid}, we give the estimated density of $e_i$ in each model, which does not strongly supports the normality assumption of three models, but all the estimated densities are close to symmetric. 
Hence, the normality assumption might be plausible.
In fact, we calculated the $p$-value of the Kolmogorov-Smirnov test for normality of $e_i$, presented in Table \ref{tab:AIC}, and found that the normality assumption was not rejected in the three models in both items.
Moreover, in Table \ref{tab:AIC}, we provide AICs based on the maximum marginal likelihood for the three models. 
It can be seen that AICs of PTFH and log-FH are similar and smaller that that of FH in ``Education" while that of PTFH is the smallest in ``Health".

\begin{figure}
\begin{center}
\includegraphics[width=6cm,clip]{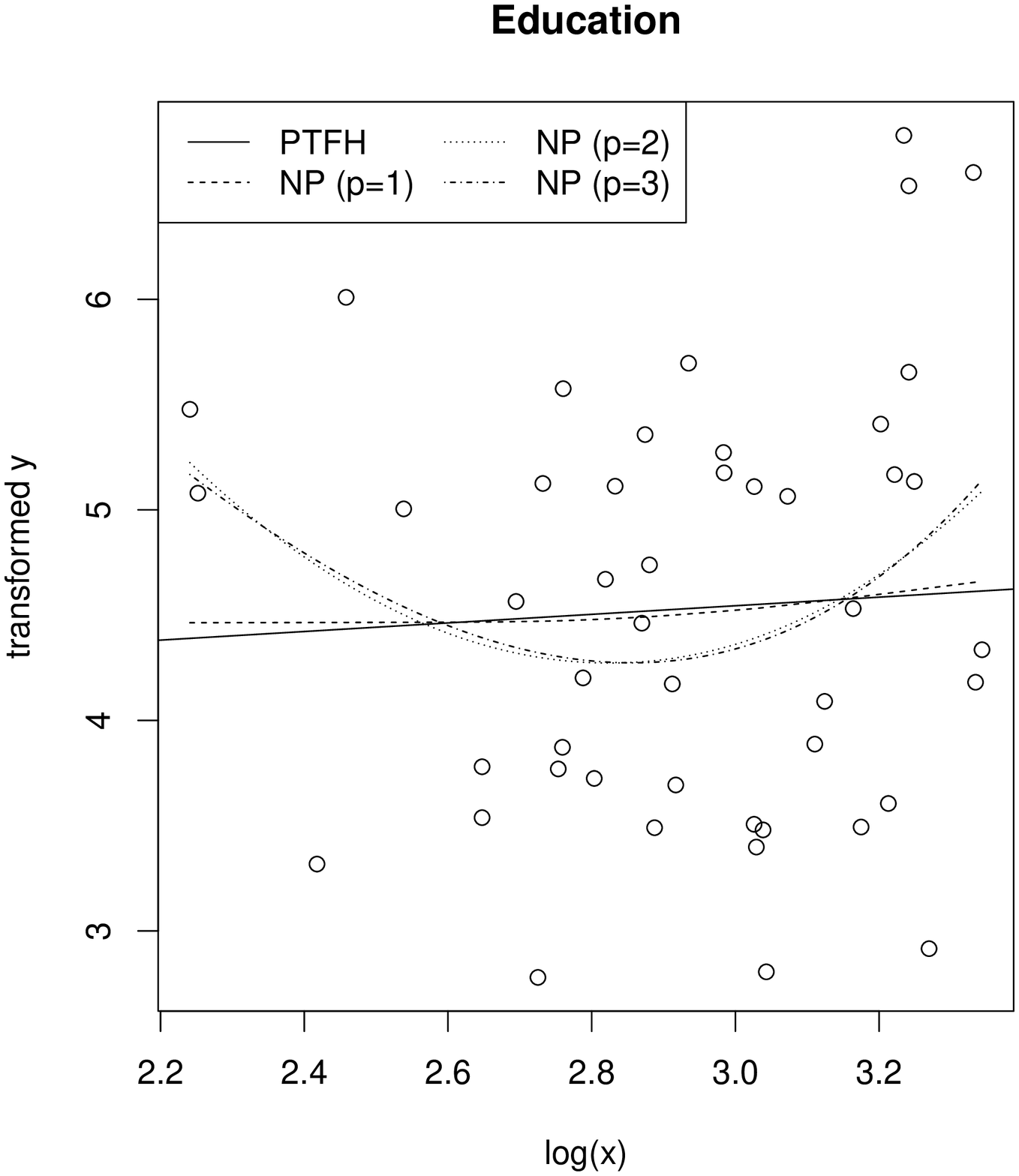}
\includegraphics[width=6cm,clip]{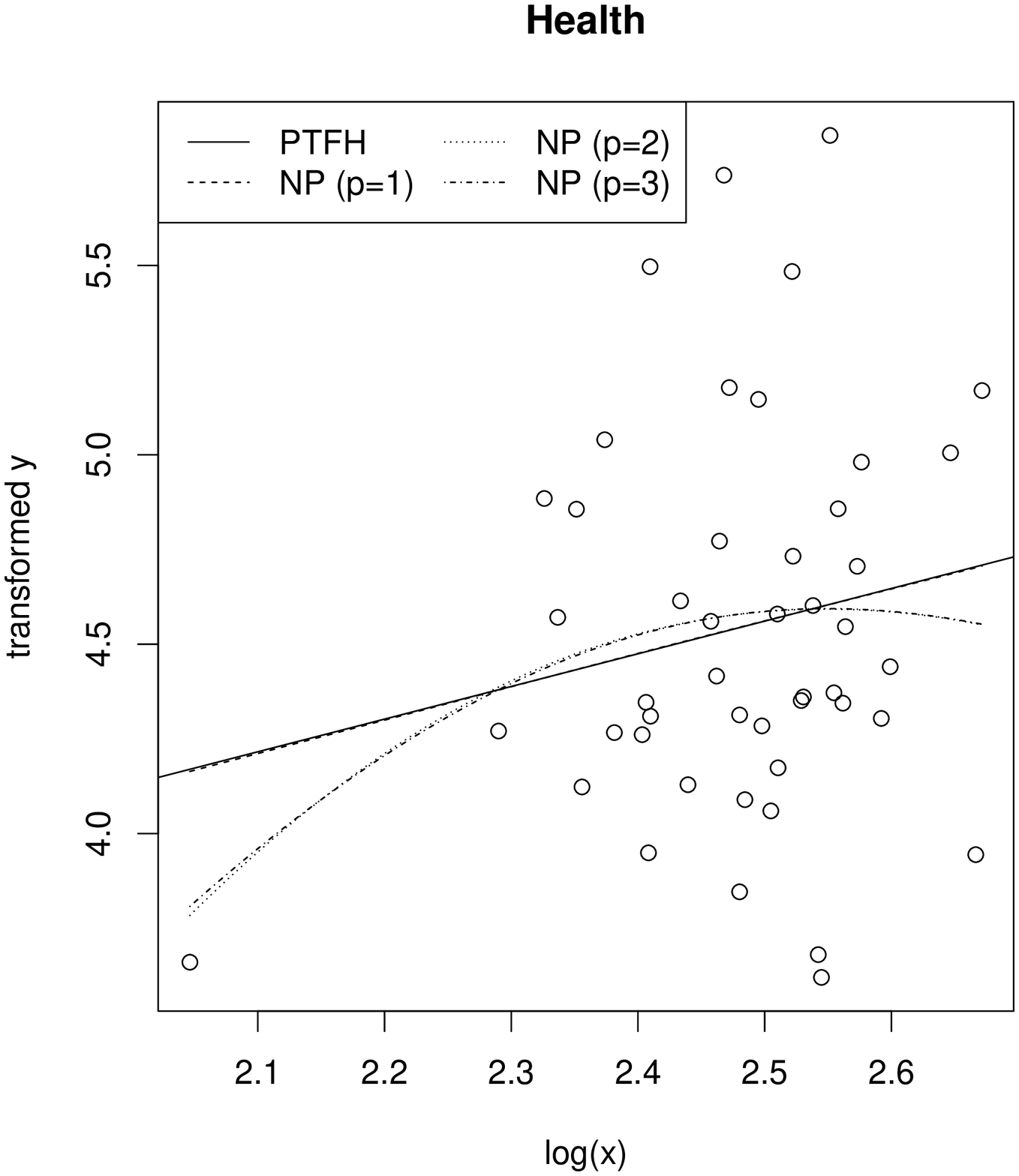}
\caption{The scatter plots of $(\log x_i,h_{\lah}(y_i))$ with estimated regression lines in the parametric transformed Fay-Herriot (PTFH) model and the nonparametric (NP) model based on the penalized spline with three orders ($p=1,2,3$).
\label{fig:app-gof}
}
\end{center}
\end{figure}

For investigation of goodness-of-fit of the PTFH model, we set $z_i=h_{\lah}(y_i)$ and $w_i=\log x_i$, and applied the penalized spline model used in Opsomer et al. (2008):
\begin{equation}\label{NP}
z_i=\be_0+\sum_{\ell=1}^p\be_{\ell}w_i^{\ell}+\sum_{\ell=1}^K\ga_{\ell}(w_i-\kappa_{\ell})^p_{+}+v_i+\ep_i, \ \ \ i=1,\ldots,m,
\end{equation}
where $v_i\sim(0,A)$, $\ep_i\sim N(0,D_i)$ and $(\ga_1,\ldots,\ga_K)^t\sim N(0,\alpha\I_K)$, $(x)_{+}^p$ denotes the function $x^pI(x>0)$, and $\kappa_1<\cdots<\kappa_K$ is a set of fixed knots which determine the flexibility of splines.
We set $K=20$ and took $\kappa_{1}$ and $\kappa_K$ as $10\%$ and $90\%$ quantiles of $w_i$, respectively, and set $\kappa_2,\ldots,\kappa_{K-1}$ at regular interval.
For the degree of splines, we considered three cases: $p=1,2,3$.
We estimated model parameters $\beta_0,\ldots,\beta_p, A$ and $\alpha$ by the maximum likelihood method.
In Figure \ref{fig:app-gof}, we present the estimated regression lines of three penalized spline models ($p=1,2,3$) as well as that of PTFH, which shows that the linear parametric structure in the PTFH model seems plausible and PTFH would fit well in both items.

Finally, we computed the MSE estimates of the small area estimators for the three models.
In the PTFH model, we used the estimator given in Theorem \ref{thm:mse} with $1000$ bootstrap samples and $5000$ Monte Carlo samples of $v_i$ and $\ep_i$ for numerical evaluation of $g_{1i}$.
For the MSE estimates in the log-FH and FH models, we used the estimator given in Slud and Maiti (2006) and Datta and Lahiri (2000), respectively.
We report the small area estimates and MSE estimates in seven prefectures around Tokyo in Table \ref{tab:app}.
It can be seen that log-FH and FH produce relatively similar estimates of area means while the estimates from PTFH are not similar to those models.
Regarding MSE estimates, we can observe that the values in PTFH are smaller than the other two models, but we cannot directly compare these results since each MSE estimates are calculated based on the different sampling variances $D_i$.

\begin{table}[!htb]
\caption{AIC and $p$-value of Kolmogorov-Smirnov (KS) test for normality of standardized residuals.
\label{tab:AIC}
}
\vspace{0cm}
\begin{center}
$
{\renewcommand\arraystretch{1.1}\small
\begin{array}{c@{\hspace{3mm}}c@{\hspace{4mm}}
c@{\hspace{3mm}}c@{\hspace{3mm}}c@{\hspace{3mm}}
c@{\hspace{2mm}}c@{\hspace{2mm}}c@{\hspace{2mm}}
c@{\hspace{2mm}}c@{\hspace{2mm}}c@{\hspace{2mm}}
c@{\hspace{2mm}}c@{\hspace{2mm}}c@{\hspace{2mm}}
}
\hline
& & \multicolumn{3}{c}{\text{AIC}} & \ \ & \multicolumn{3}{c}{p\text{-value of KS test}}\\
\text{Data} & & \text{PTFH} & \text{log-FH} & \text{FH} && \text{PTFH} & \text{log-FH} & \text{FH} \\
\hline
\text{Education} && 313.1 & 312.9 & 314.5  && 0.577 & 0.469 & 0.848 \\
\text{Health} && 172.9 & 180.7 & 183.4 && 0.519 & 0.440 & 0.375\\
 \hline
\end{array}
}
$
\end{center}
\end{table}

\begin{table}[!htb]
\caption{The small area estimates and the root of MSE (RMSE) estimates in seven prefectures around Tokyo from four models, the parametric transformed Fay-Herriot (PTFH) model, the log-transformed Fay-Herriot (log-FH) model and the classical Fay-Herriot (FH) model. 
\label{tab:app}
}
\vspace{0cm}
\begin{center}
$
{\renewcommand\arraystretch{1.1}\small
\begin{array}{c@{\hspace{3mm}}c@{\hspace{4mm}}
c@{\hspace{3mm}}c@{\hspace{3mm}}c@{\hspace{3mm}}
c@{\hspace{2mm}}c@{\hspace{2mm}}c@{\hspace{2mm}}
c@{\hspace{2mm}}c@{\hspace{2mm}}c@{\hspace{2mm}}
c@{\hspace{2mm}}c@{\hspace{2mm}}c@{\hspace{2mm}}
}
\hline
& & &&\multicolumn{3}{c}{\text{Estimates}} & \ \ & \multicolumn{3}{c}{\text{RMSE}}\\
\text{Data} &\text{Prefecture} & \text{DE} && \text{PTFH} & \text{log-FH} & \text{FH}  && \text{PTFH} & \text{log-FH} & \text{FH} \\
\hline
 & \text{Ibaraki} & 21.97 &  & 21.80 & 21.54 & 21.44 &  & 1.12 & 1.99 & 1.95 \\
 & \text{Tochigi} & 21.88 &  & 21.63 & 21.21 & 21.30 &  & 1.65 & 2.41 & 2.10 \\
 & \text{Gunma}  & 14.12 &  & 14.76 & 15.49 & 15.17 &  & 2.74 & 3.68 & 2.93 \\
\text{Education} & \text{Saitama} & 32.61 &  & 27.41 & 23.81 & 22.78 &  & 4.49 & 4.68 & 4.83 \\
 & \text{Chiba} & 21.55 &  & 20.92 & 20.19 & 20.08 &  & 3.31 & 3.90 & 3.87 \\
 & \text{Tokyo} & 22.04 &  & 21.84 & 21.55 & 21.06 &  & 1.73 & 2.16 & 3.12 \\
 & \text{Kanagawa} & 22.32 &  & 21.87 & 21.32 & 20.86 &  & 2.37 & 2.93 & 3.34 \\
\hline 
 & \text{Ibaraki} & 10.35 &  & 10.37 & 10.67 & 10.70 &  & 0.25 & 0.83 & 0.85 \\
 & \text{Tochigi} & 11.76 &  & 11.71 & 11.34 & 11.33 &  & 0.61 & 1.08 & 1.08 \\
 & \text{Gunma}  & 8.74 &  & 8.88 & 10.00 & 10.12 &  & 0.51 & 1.23 & 1.25 \\
\text{Health} & \text{Saitama} & 11.13 &  & 11.13 & 11.21 & 11.22 &  & 0.19 & 0.77 & 0.79 \\
 & \text{Chiba} & 12.81 &  & 12.64 & 11.64 & 11.64 &  & 0.60 & 1.06 & 1.07 \\
 & \text{Tokyo} & 13.80 &  & 13.73 & 12.66 & 12.53 &  & 0.18 & 0.79 & 0.85 \\
 & \text{Kanagawa} & 14.50 &  & 14.45 & 13.55 & 13.61 &  & 0.10 & 0.63 & 0.62 \\
\hline
\end{array}
}
$
\end{center}
\end{table}

\section{Concluding Remarks}\label{sec:conc}
We proposed the use of the dual power transformation in the classical Fay-Herriot model for estimating positive valued small area means, which we call parametric transformed Fay-Herriot (PTFH) model 
We derived the best predictor (BP) and obtained empirical best predictor by replacing unknown model parameters in BP with their maximum likelihood estimator.
For risk evaluation of the predictor, we derived the second unbiased estimator of mean squared errors of the predictor based on the parametric bootstrap.
In the simulation study, we have numerically evaluated the prediction errors of the proposed method together with some existing methods. 
In empirical study, we applied the method to the Japanese survey data.

For checking goodness-of-fit of PTFH, we suggested fitting the penalized spline model for the transformed data, as presented in the application.
While we did not consider estimating the penalized spline and the transformation parameter simultaneously, estimation of PTFH model with penalized spline could be useful.
However, the detailed discussion seems exceed the scope of this paper, and is left to a valuable future study.
Concerning the calculation of the sampling variance $D_i$, we only considered the case where the some auxiliary observations are available as considered in Section \ref{sec:sim} and \ref{sec:emp}.
However, due to highly variable survey weights and small samples, the sampling variance $D_i$ could be badly estimated in areas with small samples, and nothing can reassure practitioners about what effects might have on the results from the PTFH model.

Regarding model companion in Section \ref{sec:emp}, we used the AIC based on the maximum marginal likelihood (marginal AIC) for model comparison. 
However, the use of the marginal AIC in the Fay-Herriot model is known to have some drawbacks, and Han (2013) and Marhuenda et al. (2014) proposed alternative information criteria.
However, the use of these criteria in the transformation model would not be straightforward due to the variability of estimating the additional transformation parameter, thereby the detailed discussion is left to a future study.

\ \\
{\bf Acknowledgement}

\medskip
We would like to thank the Associate Editor and the four reviewers for many valuable comments and helpful suggestions which led to an improved version of this paper.
We would also like to thank professor J. N. K. Rao for helpful comments and suggestions and also thank attendees at the seminar in Statistics Canada.
This work was supported by JSPS KAKENHI Grant Numbers, JP16H07406, JP15H01943 and JP26330036.

\vspace{0.5cm}
\begin{center}
{\bf Appendix}
\end{center}

\ \\
{\bf A1. Derivation of (\ref{g1}).} \ \ \ \
Note that $\E[(\mut_i-\mu_i)^2]=\E[\mu_i^2]-\E[\mut_i^2]$.
From (\ref{BP}), it follows that
$$
\E[\mut_i^2]=\iiint_{\Re^3} h_{\la}^{-1}(s)h_{\la}^{-1}(t)\phi(s; \tht_i(u),\si_i^2)\phi(t; \tht_i(u),\si_i^2)\phi(u; \x_i^t\bbe,A+D_i)dsdtdu,
$$
where $\tht_i(u)=a_iu+(1-a_i)\x_i^t\bbe$ with $a_i=A/(A+D_i)$.
Let $S$ and $T$ be random variables mutually independently distributed as $N(\tht_i(U),\si_i^2)$ under given $U=u$, and let $U$ be a random variable distributed as $N(\x_i^t\bbe,A+D_i)$.
The marginal distribution of the vector $(S,T)'$ is 
$$
N_2\left(
\left(\begin{array}{c} \x_i^t\bbe \\ \x_i^t\bbe\end{array}\right),
A\left(\begin{array}{cc} 1 & a_i \\ a_i & 1\end{array}\right)
\right).
$$
Then, we have $\E[\mut_i^2]=\E[h_{\la}^{-1}(S)h_{\la}^{-1}(T)]$, where the expectation is taken with respect to the marginal distribution of $(S,T)'$.
Introducing random variables $z_1$ and $z_2$ mutually independently distributed as $N(0,A)$, we can express $S=\x_i^t\bbe+c_{1i}z_1+c_{2i}z_2$ and $T=\x_i^t\bbe+c_{1i}z_1-c_{2i}z_2$, thereby we obtain the expression
$$
\E[\mut_i^2]=\E[h_{\la}^{-1}(\x_i^t\bbe+c_{1i}z_1+c_{2i}z_2)h_{\la}^{-1}(\x_i^t\bbe+c_{1i}z_1-c_{2i}z_2)].
$$
Since $\E[\mu_i^2]$ can be expressed as $\E[\mu_i^2]=\E[\{h_{\la}^{-1}(\x_i^t\bbe+z_1)\}^2]$, we obtain (\ref{g1}).

\ \\
{\bf A2. Proof of Lemma \ref{lem:g2}.} \ \ \ \ 
For notational simplicity, we define
$\mut_{i(\bphi)}=\partial\mut_i/\partial\bphi$ and $\mut_{i(\bphi\bphi)}=\partial^2\mut_i/\partial\bphi\partial\bphi^t$.
Expanding $\muh_i$ around $\mut_i$, we get
$$
\muh_i-\mut_i=\mut_{i(\bphi)}^t(\bphih-\bphi)+\frac12(\bphih-\bphi)^t\mut_{i(\bphi\bphi)}(y_i;\bphi^{\ast})(\bphih-\bphi),
$$
where $\bphi^{\ast}$ is on the line connecting $\bphi$ and $\bphih$.
Then, it holds that 
$$
g_{2i}(\bphi)=\E\left[(\bphih-\bphi)^t\mut_{i(\bphi)}\mut_{i(\bphi)}^t(\bphih-\bphi)\right]+R_1+\frac14R_2,
$$
where $R_1=\E[\mut_{i(\bphi)}^t(\bphih-\bphi)(\bphih-\bphi)^t\mut_{i(\bphi\bphi)}(y_i;\bphi^{\ast})(\bphih-\bphi)]$ and $R_2=\E[\{(\bphih-\bphi)^t\mut_{i(\bphi\bphi)}(y_i;\bphi^{\ast})(\bphih-\bphi)\}]$.
We first show that $R_1=o(m^{-1})$ and $R_2=o(m^{-1})$.
We only prove $R_1=o(m^{-1})$ since the evaluation of $R_2$ is quite similar. 
In what follows, we define $\partial^2 \mut_i^{\ast}/\partial\phi_k\partial\phi_{\ell}=\partial^2 \mut_i(y_i;\bphi^{\ast})/\partial\phi_k\partial\phi_{\ell}$.
It follows that 
\begin{align*}
R_1&=\sum_{j=1}^{p+2}\sum_{k=1}^{p+2}\sum_{\ell=1}^{p+2}
E\left[\left(\frac{\partial\mut_i}{\partial\phi_j}\right)\left(\frac{\partial^2\mut_i^{\ast}}{\partial\phi_k\partial\phi_{\ell}}\right)(\phih_j-\phi_j)(\phih_k-\phi_k)(\phih_{\ell}-\phi_{\ell})\right]\\
&
\equiv \sum_{j=1}^{p+2}\sum_{k=1}^{p+2}\sum_{\ell=1}^{p+2} U_{1jk\ell},
\end{align*}
and
\begin{align*}
|U_{1jkl}|
&\leq 
\E\left[\bigg|\left(\frac{\partial\mut_i}{\partial\phi_j}\right)\left(\frac{\partial^2\mut_i^{\ast}}{\partial\phi_k\partial\phi_{\ell}}\right)\bigg|^4\right]^{\frac14} 
\E\left[\Big|(\phih_j-\phi_j)(\phih_k-\phi_k)(\phih_{\ell}-\phi_{\ell})\Big|^{\frac43}\right]^{\frac34}\\
&\leq 
\E\left[\bigg|\frac{\partial\mut_i}{\partial\phi_j}\bigg|^8\right]^{\frac18}
\E\left[\bigg|\frac{\partial^2\mut_i^{\ast}}{\partial\phi_k\partial\phi_{\ell}}\bigg|^8\right]^{\frac18} 
\prod_{a\in \{j,k,\ell\}}\E\left[\Big|\phih_a-\phi_a\Big|^{4}\right]^{\frac14}
\end{align*}
from H\"{o}lder's inequality.
From the asymptotic normality of $\bphih$ given in Lemma \ref{lem:asymp}, it follows $\E[|\phih-\phi|^r]=O(m^{-r/2})$ for arbitrary $r>0$.
Then, we have
$$
\prod_{a\in \{j,k,\ell\}}\E\left[\Big|\phih_a-\phi_a\Big|^{4}\right]^{1/4}=o(m^{-1}).
$$
Noting that 
\begin{align*}
\frac{\partial h_{\la}^{-1}(x)}{\partial \la}
&\equiv \frac{\partial}{\partial \la}\Big(\la x+\sqrt{1+\la^2x^2}\Big)^{1/\la}\\
&=\frac{h_{\la}^{-1}(x)}{\la}\left\{\frac{x}{\sqrt{1+\la^2x^2}}-\frac1{\la}\log(\la x+\sqrt{1+\la^2x^2})\right\},
\end{align*}
the straightforward calculation shows that 
\begin{align*}
\mut_{i(\la)}
&=\int_{-\infty}^{\infty}\left(\frac{\partial h_{\la}^{-1}(t)}{\partial \la}\right)\phi(t; \tht_i,\si_i^2)dt+\int_{-\infty}^{\infty}h_{\la}^{-1}(t)\left(\frac{\partial \phi(t; \tht_i,\si_i^2)}{\partial \la}\right)dt\\
&=\frac{1}{\la}\E\left[\frac{\th_ih_{\la}^{-1}(\th_i)}{\sqrt{1+\la^2\th_i^2}}\Big|y_i\right]-\frac1{\la^2}\E\left[h_{\la}^{-1}(\th_i)\log(\la \th_i+\sqrt{1+\la^2\th_i^2})\Big|y_i\right]\\
& \ \ \ \ +\frac1{D_i}\left(\frac{\partial h_{\la}(y_i)}{\partial\la}\right)\E\left[(\th_i-\tht_i)h_{\la}^{-1}(\th_i)\Big|y_i\right]\\
&\equiv \E[f_1(\th_i)|y_i]+\E[f_2(\th_i)|y_i]+\E[f_3(\th_i,y_i)|y_i],
\end{align*}
where 
$$
\frac{\partial h_{\la}(y_i)}{\partial\la}
=\frac{\log x}{\la}x^\la-\left(\log x+\frac1\la\right)h_{\la}(x).
$$
Note that 
$$
\E\left[\Big\{\E[f(\th_i,y_i)|y_i]\Big\}^a\right]
\leq 
\E\left[\E[f(\th_i,y_i)^a|y_i]\right]
=\E[f(\th_i,y_i)^a]
$$
for $a>0$ from Jensen's inequality.
Since $\E[f_1(\th_i)^a]<\infty$, $\E[f_2(\th_i)^a]<\infty$, $\E[f_3(\th_i,y_i)^a]<\infty$ for $a>0$, it follows that 
$$
\E\left[\bigg|\frac{\partial\mut_i}{\partial\la}\bigg|^8\right]<\infty.
$$
Similarly, we have
\begin{align*}
\mut_{i(\beta)}&=\frac{D_i\x_i}{(A+D_i)\si_i^2}\E\left[(\th_i-\tht_i)h_{\la}^{-1}(\th_i)\Big|y_i\right]\\
\mut_{i(A)}&=\frac{D_i^2}{2\si_i^4(A+D_i)^2}\E\left[\left\{(\th_i-\tht_i)^2-\si_i^{5/2}\right\}h_{\la}^{-1}(\th_i)\Big|y_i\right],
\end{align*}
which leads to 
$$
\E\left[\bigg|\frac{\partial\mut_i}{\partial\phi_k}\bigg|^8\right]<\infty,
$$
for $k=1,\ldots,p+1$.
Moreover, straightforward but considerable calculations shows that $\E\left[|\partial^2\mut_i^{\ast}/\partial\phi_k\partial\phi_{\ell}|^8\right] <\infty$. 
Hence, we have $R_1=o(m^{-1})$.
A quite similar evaluation shows that $R_2=o(m^{-1})$, which leads to 
$$
g_{2i}(\bphi)=\E\left[(\bphih-\bphi)^t\mut_{i(\bphi)}\mut_{i(\bphi)}^t(\bphih-\bphi)\right]+o(m^{-1}).
$$
Finally, using the similar argument given in the proof of Theorem 3 in Kubokawa et al. (2016), we have
\begin{align*}
\E\left[(\bphih-\bphi)^t\mut_{i(\bphi)}\mut_{i(\bphi)}^t(\bphih-\bphi)\right]
&=\tr\left(\E\left[\mut_{i(\bphi)}\mut_{i(\bphi)}^t\right]\E[(\bphih-\bphi)(\bphih-\bphi)^t]\right)+o(m^{-1})\\
&=\frac1m\tr\left\{\V(\bphi)\E\left[\mut_{i(\bphi)}\mut_{i(\bphi)}^t\right]\right\}+o(m^{-1}),
\end{align*}
which completes the proof.

\ \\
{\bf A3. Proof of Theorem \ref{thm:mse}.} \ \ \
Taylor series expansion of $g_{1i}(\bphih)$ around $\bphi$ gives 
\begin{align*}
\E[g_{1i}(\bphih)]=g_{1i}(\bphi)+\frac{\partial g_{1i}(\bphi)}{\partial \bphi^t}\E[\bphih-\bphi]+\frac12\tr\left(\frac{\partial^2 g_{1i}(\bphi)}{\partial\bphi\partial \bphi^t}\E[(\bphih-\bphi)(\bphih-\bphi)^t]\right)+R_3,
\end{align*}
where
$$
R_3=\frac16\sum_{j=1}^{p+2}\sum_{k=1}^{p+2}\sum_{\ell=1}^{p+2}
\frac{\partial^3 g_{1i}(\bphi)}{\partial\phi_j\partial\phi_k\partial\phi_\ell}\bigg|_{\bphi=\bphi_{\ast}}
(\phih_j-\phi_j)(\phih_k-\phi_k)(\phih_\ell-\phi_\ell).
$$
Since $\E[(\phih_j-\phi_j)(\phih_k-\phi_k)(\phih_\ell-\phi_\ell)]=o(m^{-1})$, it holds that $\E[R_3]=o(m^{-1})$.
Moreover, from Lemma \ref{lem:asymp}, we have
$$
\E[g_{1i}(\bphih)-g_{1i}(\bphi)]
=\frac1m\frac{\partial g_{1i}(\bphi)}{\partial \bphi^t}\b(\bphi)+\frac1{2m}\tr\left(\frac{\partial^2 g_{1i}(\bphi)}{\partial\bphi\partial \bphi^t}\V(\bphi)\right)+o(m^{-1}),
$$
thereby, we have $\E[g_{1i}(\bphih)-g_{1i}(\bphi)]=m^{-1}c_2(\bphi)+o(m^{-1})$ with the smooth function $c_2(\bphi)$.
Hence, from Lemma \ref{lem:g2} and Butar and Lahiri (2003), we obtain the second order unbiasedness of (\ref{estMSE})

\vspace{1cm}\noindent
{\bf References}

\end{document}